# Imaging of Alignment, Deformation and Dissociation of $CS_2$ Molecules using Ultrafast Electron Diffraction

Jie Yang, Joshua Beck, Cornelis J. Uiterwaal, Martin Centurion

University of Nebraska-Lincoln, Lincoln, Nebraska, 68588

Imaging the structure of molecules in transient excited states remains a challenge due to the extreme requirements for spatial and temporal resolution. Ultrafast electron diffraction from aligned molecules (UEDAM) provides atomic resolution and allows for the retrieval of structural information without the need to rely on theoretical models. Here we use UEDAM and femtosecond laser mass spectrometry (FLMS) to investigate the dynamics in carbon disulfide ($CS_2$) following the interaction with an intense femtosecond laser pulse. We have retrieved images of ground state and excited molecules with 0.03 Å precision. We have observed that the degree of alignment reaches an upper limit at laser intensities below the ionization threshold, and found evidence of structural deformation, dissociation, and ionization at higher laser intensities.

Laser-alignment of molecules has been proposed as a method to achieve three-dimensional diffractive imaging [1-3]. Recent experiments have shown that it is possible to record electron [4] and X-ray [5] diffraction patterns of impulsively aligned molecules, and to retrieve the molecular structure with atomic resolution [6]. Impulsive alignment is often preferable to adiabatic alignment, because the molecules can be imaged in a field free environment. The degree of alignment improves with increasing laser intensity, however, at high intensities the alignment laser pulse can cause structural deformation and ionization of the molecules. In order to determine the intensity threshold at which these changes appear, it is necessary to simultaneously determine the probability of ionization, the degree of alignment and the structure of the excited molecules. This is important to ensure that molecular imaging experiments are carried out at intensity levels that do not distort the structure. We use Ultrafast Electron Diffraction from Aligned Molecules (UEDAM) to measure the degree of alignment and structure of molecules, and Femtosecond Laser Mass Spectrometry (FLMS) to measure the degree of ionization. Carbon disulfide ($CS_2$) was chosen as a model system for this study.

The alignment, electronic excitation, and dissociation dynamics of carbon disulfide ($CS_2$) have drawn much recent interest [7-16]. Although being one of the simplest polyatomic molecules, $CS_2$ shares many general and important properties with other polyatomic molecules such as the presence of conical intersections and the ability to be photoexcited into dissociative states. Laser-based femtosecond pump-probe experiments have been a powerful method for studying molecular dynamics in the gas phase. A probe femtosecond laser pulse, for example, can ionize the molecule, and the resulting fragments (electrons and ions) studied using kinetic-energy-resolved velocity map imaging (VMI) [17, 18]. X-Ray diffraction using an X-Ray free electron laser (X-FEL) could be used to observe structural changes with Angstrom spatial resolution, provided sub-Angstrom wavelength X-Ray pulses with high fluence become available [19].

In this work UEDAM was used to investigate the dynamics of $CS_2$ after excitation with an intense ($10^{12}$-$3\times10^{13}$ W/cm$^2$) femtosecond infrared laser pulse. UEDAM allows us to retrieve both the transient molecular structure with atomic spatial resolution and the angular distribution of the molecular ensemble directly from the data without relying on theoretical modeling. FLMS was used to accurately determine the ionization products and their population as a function of intensity. Laser pulses with intensities in the range of $10^{12}$-$10^{14}$ W/cm$^2$ are often used to align molecules or to probe excited molecules by multiphoton ionization [7, 8, 12], but their influence on the molecular structure has not been fully explored. These high intensities excite a broad range of dynamical phenomena, including excitation of rotational, vibrational and electronic states leading to alignment, deformation, dissociation and ionization.

Previously, ultrafast electron diffraction (UED) has been used to measure conformational changes in molecules in the gas phase with a temporal resolution of a few picoseconds [4, 20-21], and UEDAM have been used to measure 3D static molecular structures [6]. In contrast to previous UED experiments investigating conformational changes [20], impulsive alignment allows us to obtain 2-D images of the molecular ensemble instead of only 1-D interatomic distances.

The schematic for the UEDAM experimental setup is shown in Figure 1 (a). A single laser pulse with a wavelength of 800 nm is used to excite the molecules, and a femtosecond electron pulse is used to record diffraction patterns for a range of time delays with respect to the laser pulse. The electron beam, laser beam, and gas jet are mutually orthogonal. The overall temporal resolution of the experiment is 1.0 ps, and is determined by the duration of the laser and electron pulses and the transverse width of the electron, laser and gas beams. The anisotropy of the diffraction pattern is used to monitor the degree of alignment (Figure 1 (b)). The anisotropy as a function of time is measured for each value of the laser intensity. After finding the time delay corresponding to the alignment maximum closest to zero delay, $\sim 10^9$ scattering events are accumulated to record a diffraction pattern with a high signal-to-noise ratio (SNR). The ionization products and the fraction of ionized molecules were measured using FLMS. An ultrashort laser pulse (50 fs duration, 800 nm wavelength) is focused into a dilute $CS_2$ gas ($\sim 10^{-7}$ mbar) and ion yields are recorded as a function of pulse energy using a time-of-flight ion mass spectrometer. The spectrometer has a fixed, micrometer-sized interaction volume that is smaller than the focal volume of the laser [22]. Thus, the target is a fixed number of molecules and the yields are not averaged over the spatial intensity profile of the laser focus. Further experimental details may be found in Refs. [22-25].

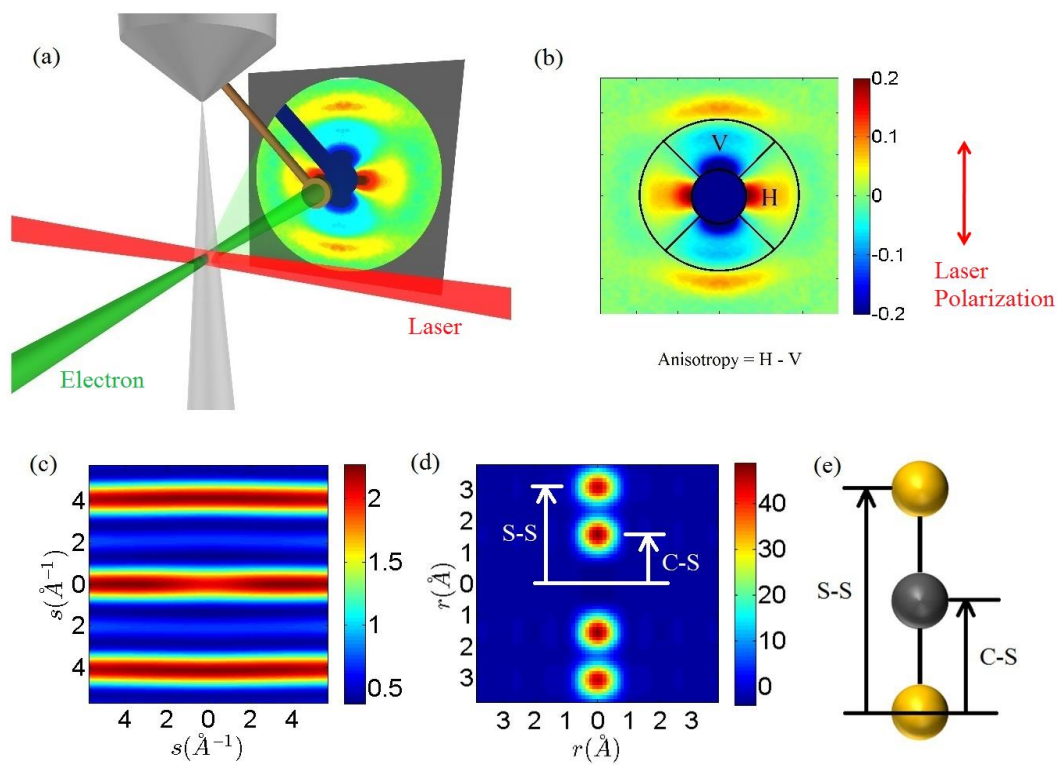

**Figure 1 | Experimental setup and data analysis methods.** (**a**) Experimental setup. The green, red and grey beams represent the electron beam, laser beam and gas jet, respectively. The beam stop is used to block the directly transmitted electrons, and the scattering pattern is recorded using a phosphor screen that is imaged on a CCD camera. (**b**) The anisotropy is calculated by taking the difference of the total counts in the horizontal (marked by "H") and the vertical (marked by "V") quadrants between two circles (more details in the Molecular images section in Methods). The diffraction patterns are normalized before calculating the anisotropy. The laser polarization is linear, shown by the red arrow. (**c**) Simulated diffraction pattern corresponding to perfect alignment. (**d**) Autocorrelation function of the molecule obtained by an inverse Fourier transform of the simulated diffraction pattern (see the Molecular images section in Methods). Each off-center spot represents an interatomic distance of $CS_2$, as shown by the two arrows. (**e**) A ball-and-stick model of $CS_2$ molecule with C-S and S-S distances marked.

The inverse Fourier transform of the diffraction pattern intensity $I$ is the autocorrelation function of the molecular structure convolved with its angular distribution (see the Molecular images section in Methods). For a linear molecule that is aligned perfectly, an image of the structure can be retrieved directly from the autocorrelation function. Figures 1 (c) and (d) show a simulation of the diffraction pattern $I$ and its inverse Fourier transform for perfectly aligned $CS_2$ molecules. The two bright spots in figure 1 (d) represent the C-S and S-S distances in the $CS_2$ molecule, as shown by the two white arrows. The spot at the center of the image, which contains no structural information, was removed for clarity. Figure 1(e) displays a model of the $CS_2$ molecule with C-S and S-S distances marked for comparison. If the alignment is not perfect, the inverse Fourier transform gives a 2-D projection of the 3-D image of the molecular ensemble, or equivalently an Abel transform of all the molecules along the direction of electron propagation.

The maximum degree of alignment that can be reached by impulsive excitation, in the limit of very short pulses, depends on the laser fluence (rather than intensity) [26]. For example, two very short pulses with different pulse durations and the same energy will produce the same alignment. Other processes, such as multiphoton excitation and ionization depend strongly on the laser intensity. We have investigated the dynamics both as a function of fluence and intensity, by using two different pulse durations (60 fs and 200 fs) for each value of the laser pulse energy. The 200 fs pulse is generated by chirping the 60 fs pulse, so both pulses have the same spectrum. We have used six different laser fluences: 0.16, 0.48, 0.79, 1.10, 1.42 and 1.73 J/cm$^2$. The laser intensity ranges from $8 \times 10^{11}$ W/cm$^2$ to $2.9 \times 10^{13}$ W/cm$^2$. Laser ionization creates a plasma that can distort the diffraction pattern due to electric and magnetic fields induced by charge separation. The strength of the plasma fields increases with time, on a timescale of several picoseconds [27]. We have carefully monitored the electron beam during experiments and found that it is not distorted, except at the point with highest intensity (1.73 J/cm$^2$ at 60 fs). This point is excluded in the following analysis, thus the diffraction patterns used in the analysis are not distorted by the plasma fields, and the maximum intensity in the UEDAM experiments is $2.4 \times 10^{13}$ W/cm$^2$. The FLMS experiments have probed higher intensities to determine the values where the sample is fully ionized.

**Results**

**Saturation of alignment and ionization.** The anisotropy was measured as a function of time for each value of the laser fluence and pulse duration. The delay between alignment laser pulse and the peak alignment decreases as the laser pulse intensity increases (this effect will be discussed in more detail later in this paper). The corresponding diffraction patterns are simulated for the experimental conditions (see the Alignment simulation and Diffraction pattern simulation sections in Methods). Only rotational excitation states are included in the simulation. A significant disagreement between measurement and simulation indicates that effects other than excitation of rotational states are present.

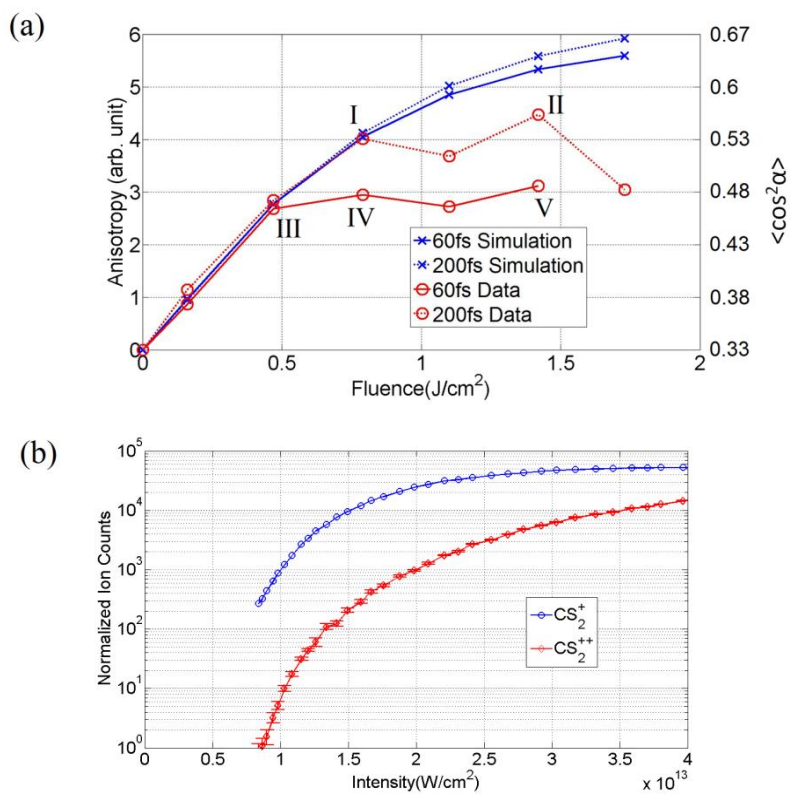

Figure 2. a) Anisotropy/alignment parameter $\langle \cos^2\alpha \rangle$ versus fluence. The blue curves are simulation results, and the red curves are experimental data. The solid and dashed curves represent pulse durations of 60 and 200 fs, respectively. The points labeled by roman numerals will be discussed below. b) Number of detected singly ionized molecules, $CS_2^+$ (blue circles), and the doubly ionized molecules $CS_2^{++}$ (red diamonds) as a function of laser intensity. The error bars for the case of the singly ionized molecule are smaller than the blue circles. To account for small pressure fluctuations, the raw counts are normalized to the average pressure for the corresponding data set.

Figure 2(a) shows the maximum anisotropy versus fluence. The peak anisotropy for each intensity is measured at the time of maximum alignment, which ranges from 1.5 ps to 0.9 ps for the 200 fs (long) laser pulses, and from 1.1 ps to 200 fs for the 60 fs (short) laser pulses. The red and blue curves in Figure 2(a) show the experimental and simulated results, respectively; while the solid lines represent 60 fs pulse durations and the dashed lines represent 200 fs pulse duration. The right axis corresponds to $\langle \cos^2\alpha \rangle$ values from simulation, where $\alpha$ represents the angle between the molecular axis and the laser polarization. The simulations show that the alignment by short or long laser pulses is very similar, with only a small disagreement in the highest intensities. The experimental results agree with the simulations at low fluence and saturate at around 0.8 J/cm$^2$ for 200 fs excitation (indicated by the roman numeral I in Figure 2) and around 0.5 J/cm$^2$ for 60 fs excitation (indicated by the roman numeral III).

This measurement indicates that dynamical processes other than excitation of rotational states are present above a threshold value of intensity that is different for different laser pulse durations. The accompanying structural changes in the molecule can be determined by examining the diffraction patterns at these points, but first we examine the role of ionization. Figure 2(b) shows the number of $CS_2^+$ ions and $CS_2^{++}$ ions as a function of laser intensity, measured using the FLMS setup. The number of $CS_2^+$ ions saturates at an intensity of $3.3 \times 10^{13}$ W/cm$^2$. At an intensity of $10^{13}$ W/cm$^2$ the number of measured ions is 0.02 times the saturation value, which means that at most 2% of the molecules have been ionized. For lower intensities, the number of ionized molecules is expected to decrease with the seventh power of the intensity (the ionization potential is 10.08 eV and the photon energy is 1.55 eV). The fraction of $CS_2^{++}$ ions remains low throughout most of the intensity range used in the UEDAM experiments: it is 0.02 % at an intensity of $10^{13}$ W/cm$^2$ and 5% at a laser intensity of $2.4 \times 10^{13}$ W/cm$^2$, which corresponds to the intensity at point V in Figure 2(a). The number of charged fragments ($S^+$ and $CS^+$) was also measured and was found to remain below 1% even at an intensity of $2.4 \times 10^{13}$ W/cm$^2$.

Figures 2(a) and 2(b) together show that the saturation of alignment occurs at much lower intensity than the saturation of ionization. For alignment with the longer laser pulses (200 fs), the alignment reaches a saturation value at an intensity of $4 \times 10^{12}$ W/cm$^2$ (point I in Figure 2(a)), where the ionization is expected to be below 1 part in $10^4$. Ionization is thus not the mechanism that prevents the alignment from further increasing with the laser fluence. All of the data points in the case of the 200 fs pulse have an intensity below $10^{13}$ W/cm$^2$. For the case of alignment with shorter laser pulses (60 fs), the alignment reaches a saturation value at a fluence of 0.48 J/cm$^2$, which corresponds to an intensity of $8 \times 10^{12}$ W/cm$^2$. Note that while the fraction of ionized molecules increase from less than a percent at point III ($8 \times 10^{12}$ W/cm$^2$) to over 60% ionized at point V ($2.4 \times 10^{13}$ W/cm$^2$), the degree of alignment remains relatively constant over this interval. While ionization could be affecting the alignment, the fact that the alignment does not decrease with increasing ionization suggests that other mechanisms are at play. For example, structural changes, which change the moment of inertia of the molecule, can have a significant impact on alignment.

**From diffraction pattern to molecular images.** We employ the diffraction-difference method: A reference pattern is subtracted from each diffraction pattern to remove unwanted signals like experimental background, scattering from the buffer gas, and scattering from unexcited molecules [6, 29]. The diffraction-difference pattern is thus due only to excited molecules. The term "diffraction-difference pattern" in this manuscript represents the difference of a diffraction pattern and a reference pattern.

Figures 3(a) and 3(b) show the diffraction-difference pattern and its inverse Fourier transform corresponding to point I in Figure 2 (a). The direction of the linear laser polarization is vertical. The reference pattern here is taken at a delay smaller than zero (the electron pulse arrives before the laser pulse), where the molecules are randomly oriented. The inverse Fourier transform in figure 3(b) at the positions corresponding to the C-S and S-S distances is positive in the vertical direction (along the laser polarization) and negative in the horizontal direction. This indicates that the number of molecules with their axis parallel to the laser polarization has increased while the number of molecules in the perpendicular direction has decreased. In other words, molecules are aligned along the laser polarization direction.

When taking randomly oriented molecules as the reference (as we do by capturing the reference pattern at negative delay), the diffraction-difference pattern reflects the angular distribution of the whole molecular ensemble. The points I and III in Figure 2 represent the fluences at which the degree of alignment saturates. By using those points as references, we can focus on molecular changes that appear for intensities beyond the saturation thresholds. We study three points after the saturation: point II for 200 fs pulse excitation and points IV and V for 60 fs excitation. These three points have slightly higher degrees of alignment than the reference points. When taking differences with these reference patterns, what remains are positive regions corresponding to molecules aligned with very narrow angular distributions and negative regions corresponding to broad angular distributions. This method circumvents the need for high alignment. Effectively, the difference gives us access to ~3% of molecules in the ensemble that are highly aligned. Structural changes can be measured with higher accuracy when the degree of alignment is high.

Figures 3(c) and 3(d) show the diffraction-difference pattern between points II and point I and the corresponding inverse Fourier transform. The laser pulse intensities at points I and II are $4\times10^{12}$ W/cm$^2$ and $7\times10^{12}$ W/cm$^2$, respectively. The difference between the two points, II-I, yields a pattern corresponding to a very narrow angular distribution that approximates perfect alignment, as shown in figure 3(c). This pattern results in an image of molecules that are highly aligned. Figure 3(d) shows a molecular image achieved using this method, where the position of the spots represents the distance between the atoms. With this method the interatomic distances can be determined with a precision of 0.03 Å. The precision is limited by maximum diffraction angle recorded in the diffraction pattern and the signal to noise ratio in the data. The precision was determined by simulating diffraction patterns under the experimental conditions, varying the interatomic distance, and running the full data processing algorithm to determine the smallest change that can be reliably detected (see Interatomic distance determination and measurement precision section in Methods).

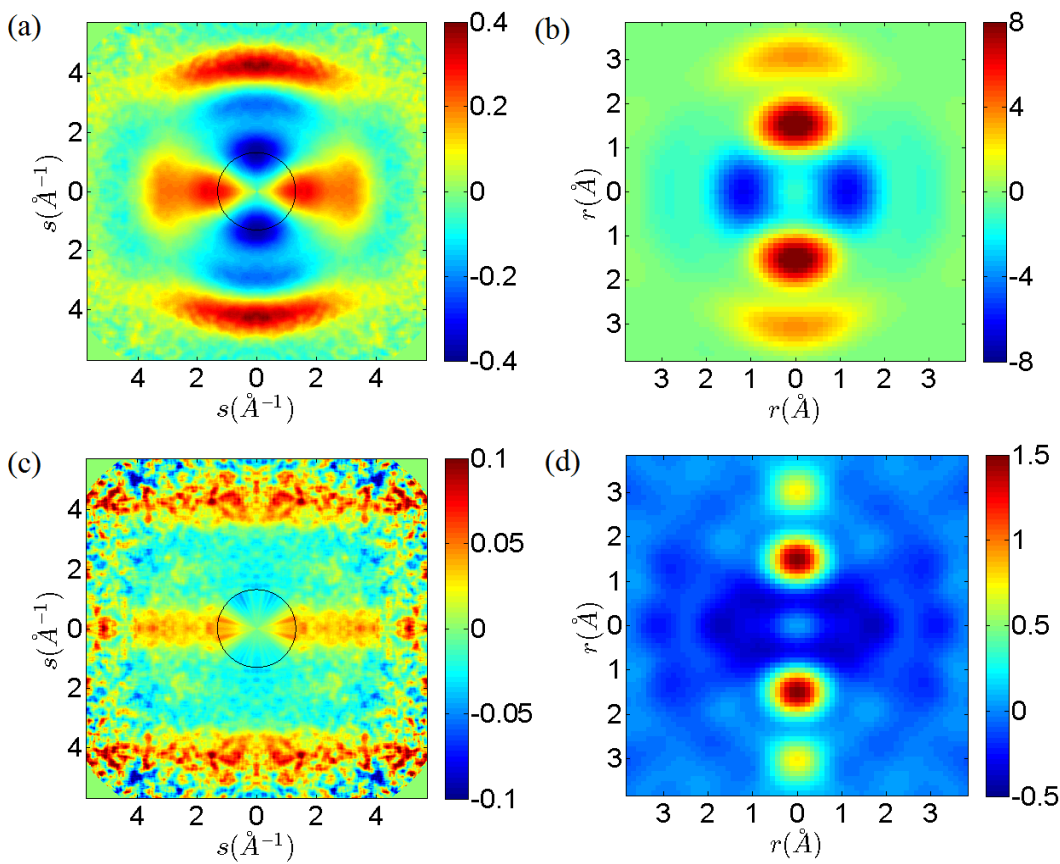

Figure 3. (a) diffraction-difference pattern of point I, reference pattern is taken at t<0; (b) inverse Fourier transform of (a); (c) diffraction-difference pattern between points II and I; (d) inverse Fourier transform of (c). The experimental data within the black circles in (a) and (c) are missing due to the beam stop, so these patterns are completed by letting the pixel values smoothly go to zero towards the center.

**Molecular image under low intensity.** Figure 4 (a) is an image of the molecule obtained from the diffraction-difference pattern between point II and point I, which corresponds to excitation with the 200 fs laser pulse. For convenience, this image is referred to as "data point II" in the following text. Figure 4(b) shows the molecular image obtained by simulating a diffraction pattern from perfectly aligned molecules in their electronic ground state. The black curves show the expected C-S and S-S distances in the ground state (1.553Å and 3.105Å, respectively). The image in Figure 4 (a) corresponds to molecules that are confined within a cone with FWHM of 9° (see Degree of alignment section in Methods). For such a narrow distribution, the molecular image is very close to that obtained with perfect alignment.

The measured bond length is consistent with that of the electronic ground state. The two interatomic distances obtained from the data are 1.53±0.03Å and 3.11±0.03Å for the C-S and S-S distances, respectively. The same data analysis performed on the simulated molecular image results in distances of 1.58±0.03Å and 3.13±0.03Å. In both cases the distances are consistent with the bond lengths in the electronic ground state. The measured distances are displayed in Table I. While the structure is consistent with the electronic ground state, the data does not rule out the excitation of vibrational modes. The excitation of vibration in addition to rotation could interfere with the alignment process, but our current experiment cannot detect small vibrations. The vibrational period would be shorter than the temporal resolution of our experiment, and the time-averaged bond distance would not change significantly. A bending vibration would result in a horizontal broadening of the CS spot. There is indeed some broadening in the CS spot in Figure 4(a), but the effect is not strong enough to draw any firm conclusions. Our method of molecular imaging is much more sensitive to changes in the position of the spots than to changes in width, so experiments with higher spatial resolution are needed to detect vibrations.

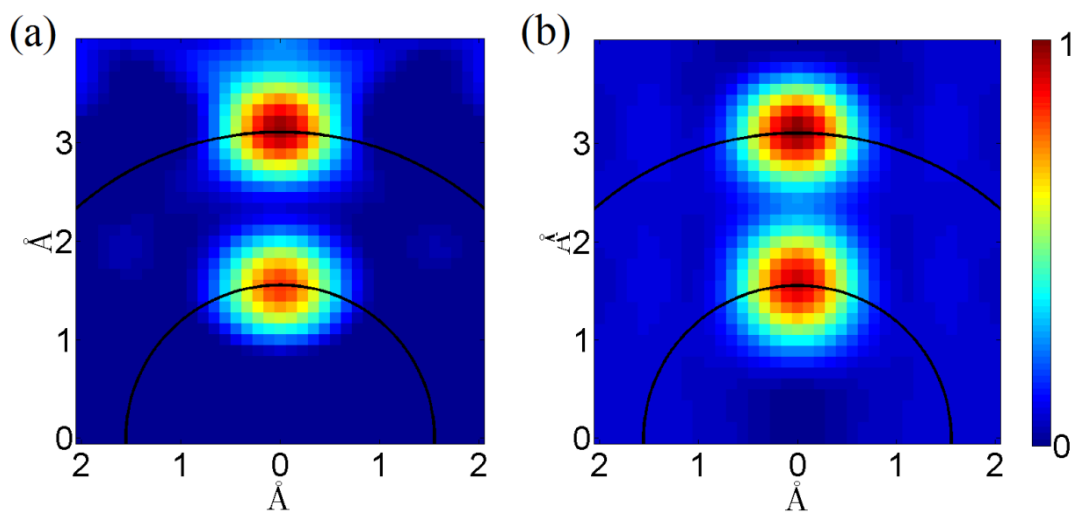

Figure 4. (a) Image of the $CS_2$ molecule obtained from the diffraction-difference pattern between points "II" ($7\times10^{12}$ W/cm$^2$) and "I" ($4\times10^{12}$ W/cm$^2$). (b) Simulation of ground state, perfectly aligned molecules. The solid black curves show the C-S and S-S distances in the ground state for reference. The images are corrected for the finite size of the electron beam on the detector (see Correction for size of the electron beam section in Methods).

**Molecular images under high intensity.** Figure 5(a) shows the image of the molecules obtained from the diffraction-difference pattern between point IV ($1.3\times10^{13}$ W/cm$^2$) and point III ($8\times10^{12}$ W/cm$^2$), which corresponds to excitation with the 60 fs laser pulse. Figure 5(b) shows the image of the molecules obtained from the diffraction-difference pattern of point V ($2.4\times10^{13}$ W/cm$^2$) with reference point III ($8\times10^{12}$ W/cm$^2$). For convenience, these two images are named as "data point IV" and "data point V" in the following text. In data points IV and V, the molecular ensemble has a lower degree of alignment than for data point II. The angular FWHM are measured to be 22° and 23°, respectively. The S-S spot is visually broader in comparison to data point II, in which a 9° angular FWHM was measured.

Both Figures 5(a) and 5(b) are significantly different from the simulated ground state (Figure 4(b)). In these two points, the S-S spots have moved upward, corresponding to a longer distance. Compared to the S-S distance in the ground state, in Figure 5(a) the distance has increased by 0.16 Å while in Figure 5(b) the distance has increased by 0.20 Å. The measurements of interatomic distances from Figure 5 are listed in Table I. For the highest rotational state populated in our simulation (J=57), the bond elongation due to centrifugal distortion is estimated to be at most $8\times10^{-4}$Å [30-31], much lower than the measured bond lengthening. The longer bond length can be attributed to electronic excitation through a multiphoton process or to the formation of ions with longer bond lengths. There are multiple excited states that can be reached by multiphoton excitation, the majority of which have longer bond lengths [32]. One example is the excitation of the $^1B_2(^1\Sigma_u^+)$ state through the absorption of 4 photons with 800 nm wavelength. In this state the S-S distance is 0.13 Å longer than in the ground state. We have also considered the possibility that the ionized molecule $CS_2^+$ would have a longer bond length. Simulations suggest that this is not the case [33], however further study of the ions is needed to completely rule out that possibility. At the highest intensity, there is a small fraction of $CS_2^{++}$ ions that might contribute to observing a longer bond length. After being excited into a dissociative state, the molecule can be ionized via subsequent multiphoton absorption, or it can dissociate into neutral fragments.

The data presents strong evidence for dissociation. In Figures 5(a) and 5(b), the ratio of the brightness of the S-S spot to the brightness of the C-S spot is reduced compared to the images in Figures 4 (a) and (b). The effect is stronger in Figure 5(b), where the laser intensity is higher. Considering a single molecule, if one of the C-S bonds breaks, the spot corresponding to the S-S distance in the molecular image vanishes as the S atom is removed, while the intensity in the C-S spot is reduced by a factor of two. Thus, when considering the whole ensemble, if a fraction of the molecules fragments, we expect that the intensity of the S-S spot would decrease more than the intensity of the C-S spot, in agreement with the measurement. Depending on the process, bond breaking can result in neutral or charged fragments. Neutral fragments are produced by electronic excitation into dissociative states, while charged fragments can be produced by bond breaking after ionization or by dissociation into neutral fragments that are then ionized by the laser. At the intensity level corresponding to point V the fraction of charged fragments was measured with the FLMS setup to be 0.7%. From the UEDAM results in Figure 5 (b), we estimate that 2/3 of the highly aligned molecules observed in the difference patterns have fragmented. Even though the difference diffraction patterns captures only a fraction of all the molecules, it is clear that the number of charged fragments is too low to account for the large fraction of fragmented molecules. Therefore, the main mechanism for bond breaking is dissociation from electronically excited neutral molecules.

As can be seen in Figures 5(a) and 5(b), and in Table I, while the distance of the S-S spot increases, the position of the C-S spot does not change and matches that of the ground state. This can be explained by considering all the contributions to the diffraction pattern and the temporal evolution of the excited states. The S-S spot results only from molecules that have not dissociated, while the C-S spot has contributions both from intact molecules and CS fragments. The CS fragment has a bond length of 1.5349Å [34], shorter than the ground state $CS_2$. The dissociation happens on a fast time scale, leading to CS+S. For example, in the case of excitation to the $^1B_2(^1\Sigma_u^+)$ state where the dissociation time has been measured, a biexponential model with two decay channels was found to be a good fit for the dissociation [7, 9]. The fast decay time constant was measured to be less than 100fs and the slow decay time was between 500fs and 800fs. Comparing this to our temporal resolution of 1.0 ps, it is clear that the measurement is time averaged over the dissociation dynamics. The diffraction patterns were recorded at the time of maximum alignment, which were 760 fs and 210 fs after excitation for points IV and V, respectively. Thus, a significant fraction of the excited molecules will have already dissociated. While further experiments with improved temporal resolution are needed to fully characterize the dynamics, our current resolution has allowed us to capture a snapshot of the structure of the excited state.

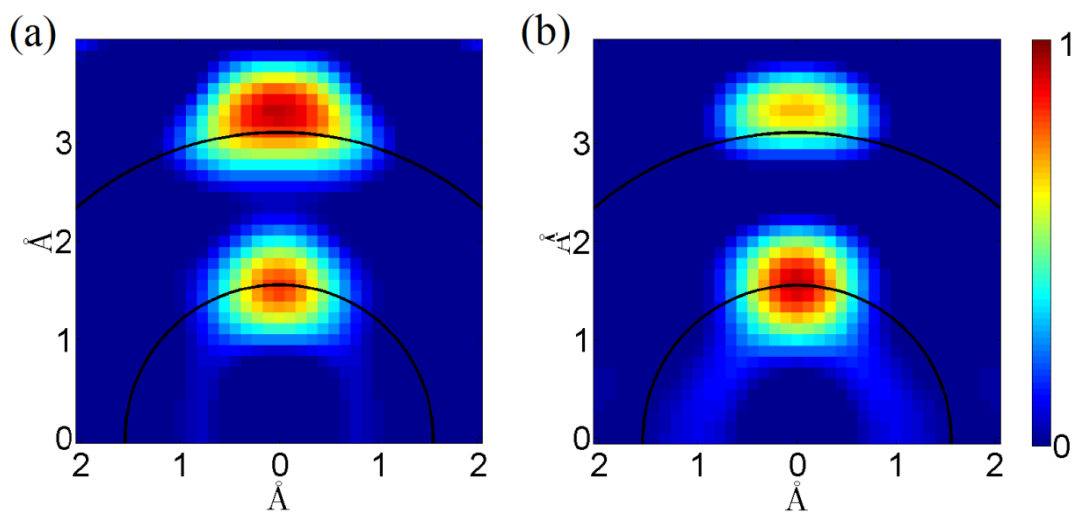

Figure 5. (a) Image of $CS_2$ molecules taken from the diffraction-difference pattern between points "IV" ($1.3 \times 10^{13}$ W/cm$^2$) and "III" ($8 \times 10^{12}$ W/cm$^2$). (b) Image of $CS_2$ molecules taken from the diffraction-difference pattern between points "V" ($2.4 \times 10^{13}$ W/cm$^2$) and "III" ($8 \times 10^{12}$ W/cm$^2$).

|  | C-S Distance (Å) | S-S Distance (Å) |
| --- | --- | --- |
| Expected Interatomic Distances for the Ground State | 1.553 | 3.105 |
| Ground State Simulation | 1.58±0.03 | 3.13±0.03 |
| Data Point "II" | 1.53±0.03 | 3.11±0.03 |
| Data Point "IV" | 1.52±0.03 | 3.27±0.03 |
| Data Point "V" | 1.55±0.03 | 3.31±0.03 |

Table I. Expected interatomic distances for ground state and interatomic distances

extracted from simulated and measured diffraction patterns. The uncertainty in the interatomic distance is explained in the Methods section Interatomic distance determination and measurement precision.

Our interpretation of the data is that for the alignment experiments with 200 fs laser pulses the saturation of alignment is most likely caused by excitation of vibrational modes, since at this intensity level ionization is not significant and no structural changes were observed. For the measurements with 60 fs pulses, where the intensity is higher, the saturation of alignment is due to different mechanisms. Bond lengthening, dissociation, and ionization are observed in this intensity range. We cannot separate the effect of ionization and structural changes on the alignment, but the fact that the degree of alignment remains relatively constant even as the degree of ionization increases by two orders of magnitude suggests that ionization is not the limiting mechanism. In the next section, we present additional evidence from the rotational dynamics that supports the previous evidence for dissociation.

**Timing of the first alignment peak.** The time delay for the molecules to reach the maximum degree of alignment after interacting with the laser pulse depends strongly on the moment of inertia of the molecule, as well as the laser pulse intensity. The peak alignment is defined as the maximum anisotropy in the pattern, as shown in Figure 1(b). If the $CS_2$ molecule dissociates, the CS fragment, which has a lower moment of inertia, will reach its maximum alignment sooner. Figure 6 shows the time delay between the laser pulse and the first alignment peak as a function of fluence for (a) the 60 fs laser pulse and (b) the 200 fs laser pulse. In the case of the 200 fs laser pulse, the time delay slowly decreases with laser intensity, in good agreement with theory. This is also in good agreement with our previous conclusion that for the case of the longer pulses there are no significant structural changes (small vibrations would not significantly change the moment of inertia). For 60 fs pulses where higher intensities are reached, experiment and theory agree for the lower fluence values but start to differ at a fluence of 0.7 J/cm$^2$ (intensity $1.3 \times 10^{13}$ W/cm$^2$), which corresponds to data point IV. This is the first point where we expect dissociation, and it also agrees with the saturation of the alignment in Figure 2. The maximum alignment has contributions from all the molecules, only a fraction of which are dissociated, therefore the moment of inertia cannot be directly retrieved from the measured time delay. However, the fact that the delay becomes significantly shorter than expected supports the conclusion that dissociation is significant at an intensity of $1.3 \times 10^{13}$ W/cm$^2$.

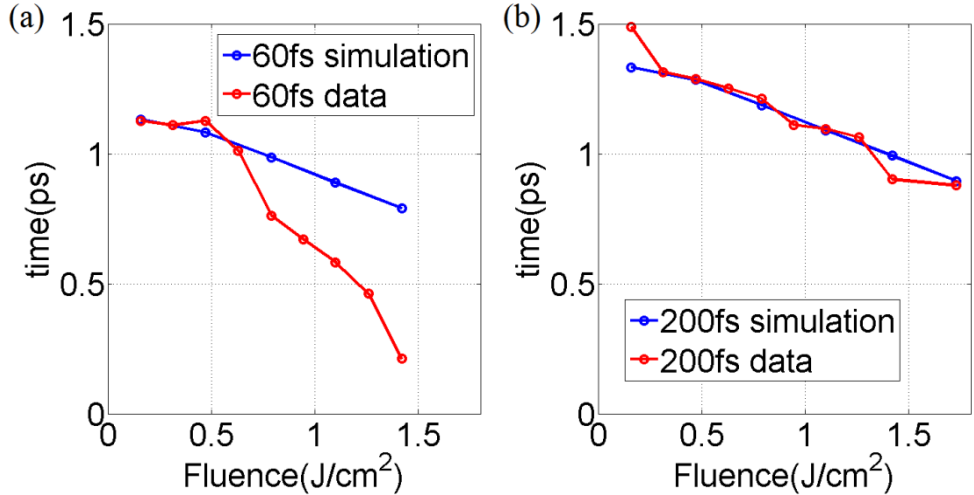

Figure 6. Time delay between the laser and the peak alignment for (a) 60 fs and (b) 200 fs. The blue and red curves represent simulated and experimental data, respectively.

**Discussion**

We have investigated the interaction of intense laser pulses with $CS_2$ molecules and have observed saturation of alignment, deformation, dissociation and ionization. For excitation with long (200 fs) laser pulses, no structural changes are observed. The saturation of the alignment is most likely caused by the excitation of vibrational states. For higher intensities, multiphoton excitation leads to deformation, dissociation and ionization of the molecule. This electronic excitation also leads to saturation of the degree of alignment. Our conclusions are supported by spatially resolved molecular images, by the dynamics of the alignment process, and by independent measurements of the degree of ionization. Our results provide clear guidelines on intensity levels that can be used for impulsive alignment experiments without introducing structural changes, and show that the degree of alignment may reach a maximum value well below the threshold for multiphoton ionization. While the alignment saturates, there appears to be a range of intensities over which the structure does not change significantly, which can be a window for molecular imaging experiments. An alternative approach could be to use adiabatic alignment, where a long laser pulse is used to align the molecules, and the diffraction pattern is captured in the presence of the laser field [35]. Using a temporally shaped alignment laser pulse with slow ramp-up and rapid turn off [28] can in principle achieve both higher degree of alignment and a field-free environment for diffraction. A similar study of electron diffraction under adiabatic alignment conditions would shed light on the intensity limits that can be used in that method and any additional effects that might result from performing the diffraction in the presence of a strong laser field. For future experiments, the temporal resolution of UEDAM could be improved to around 100 fs using keV electron pulses compressed by an RF cavity [36] in combination with tilted laser pulses [37-38] or by using MeV electron beams [39]. The spatial resolution can be improved by increasing the beam current, which will allow increasing the area of momentum space captured in the diffraction patterns.

**Methods**

**UEDAM experiment.** $CS_2$ molecules are seeded into buffer gas of Helium with a 1:20 ratio for rotational cooling. The de Laval nozzle has a 30 μm diameter at the throat and a 90 μm diameter at the exit. A supersonic jet is formed after the nozzle. The nozzle backing pressure is set to 1000 Torr. The sizes of electron beam, laser beam and gas jet are 100μm, 100μm and 200μm FWHM, respectively. The $CS_2$ gas density is ~$5\times10^{15}$/cm$^3$ in the interaction region, and the rotational temperature is estimated to be around 30 K [40]. The electron pulses are generated using a photocathode and accelerated in a static field to a kinetic energy of 25 keV, which corresponds to $\lambda_e = 7.7\ pm$. The number of electrons per pulse is set to 1000 to reduce the space-charge effect and maintain a pulse duration of approximately 400 fs at the sample. The overall instrument response function for this experiment, including the velocity mismatch between photons and electrons, is 1.0 ps. This is sufficient to capture diffraction patterns since the lifetime of the alignment is approximately 2 ps. The laser generates pulses with 2 mJ energy and a wavelength of 800 nm at a repetition rate of 5 kHz.

**Molecular images.** All diffraction patterns are normalized to the scattering from individual atoms: $I = I_{raw}/I_{at}$, where $I$ represents the normalized diffraction pattern, $I_{raw}$ is the raw diffraction pattern, and $I_{at}$ is the sum of the scattering intensity of individual atoms.

In this work, anisotropy is used to trace the evolution of alignment. Anisotropy is defined as the difference in total counts between the horizontal and the vertical quadrants in the range of $s$ = 1-3.5 Å$^{-1}$ (figure 1 (b)) in the normalized diffraction pattern. The quantity $s = \frac{4\pi}{\lambda_e}sin(\theta/2)$ is the momentum change of the scattered electrons, where $\lambda_e$ is the electron wavelength and $\theta$ is the scattering angle.

The inverse Fourier transform of the diffraction pattern intensity (without phase) of a single molecule gives the autocorrelation function of the molecular structure. A diffraction pattern of a molecular ensemble is an incoherent sum of diffraction patterns from each molecule:

$$\mathcal{F}^{-1}\left[\sum_i |\mathcal{F}[a_i]|^2\right] = \sum_i \mathcal{F}^{-1}[|\mathcal{F}[a_i]|^2] = \sum_i a_i * a_i$$

In this equation, the left side is an inverse Fourier transform of the incoherent sum of many single-molecule diffraction patterns, and the right side is the sum of the autocorrelation functions of all molecules.

**Correction for the size of the electron beam.** The electron beam is focused onto the gas jet, and the size of the beam increases as it propagates after the gas jet. Thus, the finite size of the electron beam on the detector will reduce the sharpness of diffraction patterns. The diffraction pattern measured experimentally is a convolution of the ideal diffraction pattern with the measured 2D shape of the electron beam on the detector. This is equivalent to multiplying the autocorrelation functions by the inverse Fourier transform of the electron beam. Therefore, this effect is corrected by dividing the autocorrelation functions by an envelope that is the inverse Fourier transform of the electron beam.

**Interatomic distance determination and measurement precision.** The interatomic distances are determined by the Gaussian approximation sub-pixel peak position determination algorithm [41]. The 0.03Å precision is calculated by applying this algorithm to simulated patterns with 50 different bond lengths and taking the standard deviation of the difference between output lengths and input lengths. In our experiment, the spatial resolution is given by $\delta=2\pi/s_{max}\approx1.2$Å, which gives the width of the measured spots in the molecular images. The precision in the interatomic distances is given by the position of the spots, which can be determined to higher accuracy than the width of the spots, as long as they are well separated. As a comparison, static gas-phase electron diffraction experiments [42] typically achieve a precision of 0.001 Å with a resolution of $\delta\approx0.2$Å.

**Degree of alignment.** The degree of alignment is usually measured by the quantity $\langle\cos^2\alpha\rangle$. However, this metric is not applicable to the case of difference-diffraction patterns. In this case the total population will sum up to zero. The molecular images in this work contain a narrow positive population section and a broad negative population section. The positive population constructs a cone shape. We use the full width at half maximum (FWHM) of the full-opening angle of the positive cone to characterize the degree of alignment. The angular distribution is extracted using the pBasex Abel inversion algorithm [43]. The finite scattering angle captured in the diffraction patterns, corresponding to a convolution with a Gaussian envelope in the molecular image, is taken into account in the algorithm.

**Alignment simulation.** The simulation of impulsive alignment is calculated using a linear rigid rotor interacting with a non-resonant pulse described by the time-dependent Schrodinger equation [44]. In order to better match experimental conditions, we have also simulated the effect of a pre-pulse and a pedestal with up to 1% of the peak intensity and 20 times longer pulse duration. Neither of these significantly changed the maximum degree of alignment.

**Diffraction pattern simulation.** The simulations of diffraction patterns are calculated using cylindrical harmonics [45]. In the simulations, the atoms are assumed to be stationary at their equilibrium positions without vibrations.

**Acknowledgement**

J. Yang and M. Centurion were supported by the U.S. Department of Energy Office of Science, Office of Basic Energy Sciences under Award Number DE-SC0003931. J. Beck and C. J. Uiterwaal were partially supported by U.S. Department of Education GAANN grants numbers P200A090156 and P200A120188 and by the National Science Foundation EPSCoR RII Track-2 CA Award No. IIA-1430519 (Cooperative Nebraska-Kansas grant). We would like to thank Christopher Hensley for his contribution to the initial setup of this experiment.

**Contributions**

M.C. designed the UEDAM experiments; J.Y. performed the UEDAM experiments; M.C. and J.Y. developed data analysis methods; J.Y. analyzed UEDAM experimental data; J.Y. simulated alignment process and diffraction patterns; J.B. performed the

FLMS measurements, J.B. and C.J.U. analyzed the FLMS data. M.C. and J.Y. wrote the manuscript; M.C. supervised the project.

**Competing financial interests**
The authors declare no competing financial interests.